\documentclass[twocolumn]{aastex63}

\usepackage{mathrsfs}

\shorttitle{JWST spectroscopy of candidate obscured quasars at $z \sim 6$}
\shortauthors{Matsuoka et al.}

\begin{document}

\title{Aether-SHELLQs: JWST integral-field spectroscopy of candidate obscured quasars at $z \sim 6$}

\correspondingauthor{Yoshiki Matsuoka}
\email{matsuoka.yoshiki.ld@ehime-u.ac.jp}


\author[0000-0001-5063-0340]{Yoshiki Matsuoka}
\affil{Research Center for Space and Cosmic Evolution, Ehime University, Matsuyama, Ehime 790-8577, Japan.}

\author[0000-0002-2662-8803]{Roberto Decarli}
\affil{INAF -- Osservatorio di Astrofisica e Scienza dello Spazio di Bologna, via Gobetti 93/3, I-40129 Bologna, Italy.}

\author[0000-0002-6822-2254]{Emanuele Paolo Farina}
\affil{Gemini Observatory, NSF's NOIRLab, 670 N A`ohoku Place, Hilo, HI 96720, USA.}

\author[0009-0009-8274-441X]{Anniek J. Gloudemans}
\affil{Gemini Observatory, NSF's NOIRLab, 670 N A`ohoku Place, Hilo, HI 96720, USA.}



\author[0000-0002-2931-7824]{Eduardo Ba\~nados}
\affil{Max Planck Institut f\"ur Astronomie, K\"onigstuhl 17, D-69117 Heidelberg, Germany.}

\author[0000-0002-4770-6137]{Fabrizio Arrigoni Battaia}
\affil{Max-Planck-Institut f\"ur Astrophysik, Karl-Schwarzschild-Stra\ss e 1, D-85748 Garching bei M\"unchen, Germany}

\author[0000-0003-2895-6218]{Anna-Christina Eilers}
\affil{Department of Physics, Massachusetts Institute of Technology, Cambridge, MA 02139, USA}
\affil{MIT Kavli Institute for Astrophysics and Space Research, Massachusetts Institute of Technology, Cambridge, MA 02139, USA.}

\author[0000-0002-5941-5214]{Chiara Mazzucchelli}
\affil{Instituto de Estudios Astrof\'{\i}sicos, Facultad de Ingenier\'{\i}a y Ciencias, Universidad Diego Portales, Avenida Ej\'{e}rcito Libertador 441, Santiago, Chile.}

\author[0000-0002-0106-7755]{Michael A. Strauss}
\affil{Department of Astrophysical Sciences, Princeton University, Peyton Hall, Princeton, NJ 08544, USA.}

\author[0000-0002-2536-1633]{Hyewon Suh}
\affil{Gemini Observatory, NSF's NOIRLab, 670 N A`ohoku Place, Hilo, HI 96720, USA.}

\author[0000-0002-6849-5375]{Maxime Trebitsch}
\affil{Observatoire de Paris, PSL University, Sorbonne Universit\'e, CNRS, 75014 Paris, France.}

\author[0000-0003-4793-7880]{Fabian Walter}
\affil{Max Planck Institut f\"ur Astronomie, K\"onigstuhl 17, D-69117 Heidelberg, Germany.}

\author[0000-0002-7633-431X]{Feige Wang}
\affil{Department of Astronomy, University of Michigan, 1085 S. University Ave., Ann Arbor, MI 48109, USA.}

\author[0000-0003-4569-1098]{Kentaro Aoki}
\affil{Subaru Telescope, National Astronomical Observatory of Japan, Hilo, HI 96720, USA.}

\author[0009-0007-0864-7094]{Junya Arita}
\affiliation{Department of Astronomy, School of Science, The University of Tokyo, Tokyo 113-0033, Japan.}

\begin{abstract}

We present James Webb Space Telescope (JWST) NIRSpec integral field unit (IFU) observations of six galaxies at $z\sim6$, obtained as part of the {\em Aether} project
(General Observers program 5645).
The targets were originally identified  by the Subaru High-$z$ Exploration of Low-Luminosity Quasars (SHELLQs) survey,
as candidate obscured quasars with luminous ($\gtrsim10^{43}$ erg s$^{-1}$) but narrow ($\lesssim500$ km s$^{-1}$) Ly$\alpha$ emission.
Two objects exhibit a broad component in their Balmer lines (FWHM $>3000$ km s$^{-1}$), indicating the presence of active galactic nuclei (AGNs), while the remaining four show similar profiles in permitted and forbidden lines.
Combining these data with similar SHELLQs objects reported previously, we find that the presence of broad lines is strongly correlated with Ly$\alpha$ luminosity ($L_{\rm Ly\alpha}$); 
the inferred AGN fraction is $>$77 \% and $<$15 \% above and below $L_{\rm Ly\alpha} =10^{44}$ erg s$^{-1}$, respectively.
Dust-extinction corrections inferred from the Balmer decrement would imply unrealistically high Ly$\alpha$ luminosities, suggesting that the line-emitting gas consists of multiple zones.
The IFU data reveal diverse spatial structures.
The AGN hosts are compact, whereas the other galaxies show extended ionized gas on scales up to 10 kpc and star formation rates of 60 -- 600 $M_\odot$ yr$^{-1}$.
One of the extended objects exhibits a signature of rotation, while most of the others show little ordered kinematics, with velocity widths (FWHM) up to 200 -- 300 km s$^{-1}$.
These objects populate the intermediate luminosity regime between classical luminous quasars and the low-luminosity AGNs discovered by JWST, including Little Red Dots, potentially linking the two populations.
\clearpage
\end{abstract}




\section{Introduction} \label{sec:intro}

Observations over the past few decades have revealed an unexpectedly complex population of active galactic nuclei (AGNs) in the epoch of reionization (EoR, referring to $z \ge 6$ in this paper).
On one hand, wide-field ($>$100 deg$^2$) surveys with ground-based facilities discovered classical luminous quasars, which have typical absolute magnitudes of $M_{\rm UV} < -24$ 
at the rest-frame ultraviolet (UV) wavelengths \citep[e.g.,][]{fan23}.
These quasars exhibit the familiar signatures of AGN activity, including strong X-ray and mid-infrared emission, radio emission in roughly 10\% of sources
\citep[e.g.,][]{banados15}, together with flux variability.

On the other hand, the James Webb Space Telescope \citep[JWST;][]{rigby23} has uncovered a numerous population of galaxies showing broad components in \ion{H}{1} Balmer lines,
with the full-widths at half maximum (FWHM) exceeding 1000 km s$^{-1}$ \citep[e.g.,][]{carnall23, harikane23,  larson23, onoue23, kokorev23, ubler23, fujimoto24, furtak24, juodzbalis24, juodzbalis25,killi24,labbe24, lin24, lin25, kocevski23,kocevski24,  maiolino24, schindler24, taylor24, taylor25, tripodi24, wang24b, akins25, naidu25_lrd}.
These objects have been found from observations covering a few 100 arcmin$^2$ at most, and are characterized by much lower luminosities with $M_{\rm UV} > -21$.
Many of them are associated with the population called ``Little Red Dots" \citep[LRDs;][]{matthee24}, defined by compact morphology and V-shaped spectral energy distributions (SEDs)
with blue rest-UV and red rest-optical colors \citep[e.g.,][]{labbe23, labbe25, setton24}.
A number of theoretical studies suggest that LRDs are mass-accreting supermassive black holes (SMBHs) surrounded by optically-thick gas envelop 
\citep[e.g.,][]{inayoshi25, liu25, lin26, rusakov26}, but this scenario has yet to be firmly established.

If these galaxies are interpreted as low-luminosity AGNs, their number density is surprisingly high. 
It exceeds expectations from extrapolations of the classical quasar luminosity function \citep[LF; e.g.,][]{p5, p19, niida20, schindler23} by one to two orders of magnitude 
\citep[e.g.,][]{harikane23, greene24, kocevski24, maiolino24}.
However, unlike classical quasars, these objects lack X-ray  \citep{ananna24, maiolino24_xray, yue24}, 
mid-infrared  \citep[e.g.,][]{wang24,williams24,setton25}, 
and radio detections \citep[e.g.,][]{akins24, mazzolari24, gloudemans25, perger25}, show little variability \citep{kokubo24, tee24}, 
and exhibit SEDs that are difficult to reproduce with standard AGN-galaxy composite models  \citep[][but see, e.g., \citealp{labbe24}]{ma25}.
Thus, two apparently distinct AGN populations now coexist in the EoR, separated not only by luminosity but also by their multi-wavelength properties.
The number density of LRDs 
has been reported to peak around the EoR \citep[e.g.,][]{ma25b}, but more recent studies suggest that the redshift distribution may be more extended,
reaching a maximum around 
$z \sim 5 - 6$, once selection biases are taken into account \citep[e.g.,][]{rinaldi26}.
On the other hand, quasars become much more abundant at lower redshifts.

A crucial question is whether, and how, these populations are physically connected.
The answer may lie in the intermediate luminosity regime ($-24 < M_{\rm UV} < -21$), which remains poorly explored.
Objects in this range are too rare for deep JWST pencil-beam surveys, yet too faint to be efficiently identified in most past wide-field surveys.
Without a systematic study of this regime, it remains unclear whether the apparent dichotomy reflects a fundamentally different nature, such as accretion modes of SMBHs or surrounding gas environments
\citep[e.g.,][]{inayoshi24, pacucci24, inayoshi25}. 

The Subaru High-$z$ Exploration of Low-Luminosity Quasars \citep[SHELLQs;][]{p1} survey is uniquely suited to address this gap.
It is a project to search for EoR quasars based on the Hyper Suprime Cam \citep[HSC;][]{miyazaki18} Subaru Strategic Program \citep{aihara18} survey, 
whose widest component has imaged an area of 1100 deg$^2$ down to the 5$\sigma$ limiting magnitudes of ($g$, $r$, $i$, $z$, $y$) = (26.5, 26.5, 26.2, 25.2, 24.4) for point sources \citep{aihara22}.
SHELLQs has so far reported spectroscopic identification of 182 unobscured quasars at $5.6 < z < 7.1$ \citep{p2,p4,p7,p10,p16,p20,p24}.
This sample extends well into the intermediate luminosity range.
Extensive follow-up observations are being carried out with JWST, Atacama Large Millimeter/submillimeter Array (ALMA), and other facilities, tracing the stellar 
\citep[][M. Onoue et al. 2026, in preparation]{ding23, ding25, onoue24} and
 gaseous \citep{izumi18, izumi19, izumi21a,izumi21b, phillips25, sawamura25} components of the host galaxies as well as the surrounding environments \citep{arita26}.

The SHELLQs survey has also identified 35 galaxies with extremely strong Ly$\alpha$ emission at $5.8 < z < 6.9$.
Based on the high luminosities ($L_{\rm Ly\alpha} > 10^{43}$ erg s$^{-1}$) and small widths (FWHM $<$ 500 km s$^{-1}$), these objects have been reported as candidate obscured (narrow-line) quasars.
\citet{cy2} observed 11 of them (and two quasars with similar Ly$\alpha$ profiles) with JWST/NIRSpec, and found a surprisingly high detection rate ($>$60 \%) of
broad Balmer lines, confirming the presence of AGNs.
The Balmer decrements indicate dust obscuration of up to $A_V \sim 3$, which is likely responsible for the absence of broad Ly$\alpha$ in the rest-UV spectra.  
Some of these objects show continuum slopes that are consistent with the V-shape spectra of LRDs.
Furthermore, the non-detection in Chandra observations \citep{iwasawa25} indicates X-ray luminosities much lower than expected from the rest-optical emission assuming a typical AGN SED, similar to what is seen in LRDs.
These partially-obscured AGNs with intermediate luminosities may thus represent a key population, bridging classical UV-luminous quasars and the low-luminosity AGNs identified by JWST.

This paper presents JWST/NIRSpec integral field unit (IFU) observations of six SHELLQs objects, drawn from the above class of candidate obscured quasars.
By examining both nuclear and extended emission properties, we aim to investigate the nature of these intriguing objects, including the presence of AGN signatures and the global gas kinematics.
We describe the observations and data reduction in \S \ref{sec:obs}, and then present results and discussion in \S \ref{sec:results}.
A summary and conclusions follow in \S \ref{sec:summary}.
We adopt the cosmological parameters $H_0$ = 70 km s$^{-1}$ Mpc$^{-1}$, $\Omega_{\rm M}$ = 0.3, and $\Omega_{\rm \Lambda}$ = 0.7.
All magnitudes are presented in the AB system \citep{oke83}. 
The rest-UV absolute magnitudes ($M_{\rm UV}$) and Ly$\alpha$ line luminosities ($L_{\rm Ly\alpha}$) and widths are reported as measured, i.e., without correction for dust extinction or intergalactic medium (IGM) absorption, unless otherwise noted.
We report wavelengths in both the rest frame ($\lambda_{\rm rest}$) and the observed frame ($\lambda_{\rm obs}$), depending on the context.

\section{Observations} \label{sec:obs}

\begin{deluxetable*}{cccccc}
\tablecaption{Targets of this work\label{tab:targets}}
\tablewidth{0pt}
\tablehead{
\colhead{Name (Formal HSC name)} & \colhead{$z_{\rm Ly\alpha}$} & \colhead{$\log\ (L_{\rm Ly\alpha}/{\rm [erg\ s}^{-1}])$} 
& \colhead{FWHM (km s$^{-1}$)} & \colhead{$M_{\rm UV}$} & \colhead{Ref}
}
\startdata
G03 ($J$093543.32$-$011033.3)   & 6.08 & 44.12  $\pm$ 0.01 & $<$230  &  $-$21.97  $\pm$ 0.18 & 4 \\ 
G04 ($J$125437.08$-$001410.7)   & 6.13  & 44.03  $\pm$ 0.01 & 310 $\pm$ 20  &  $-$20.91  $\pm$ 0.32 & 4\\  
G08 ($J$090544.65$+$030058.9)   & 6.27  & 43.89  $\pm$ 0.02 & 250 $\pm$ 40  &  $-$22.55  $\pm$ 0.11 & 2 \\ 
G12 ($J$120924.01$-$000646.5)    & 5.86 & 43.01 $\pm$ 0.04 &  580 $\pm$ 50 &  $-$22.51 $\pm$ 0.05 & 3 \\
G13 ($J$120754.14$-$000553.3)    & 6.01 &  42.92 $\pm$ 0.05 &  420 $\pm$ 160 &  $-$22.77 $\pm$ 0.06  & 1\\
G14 ($J$121503.42$-$014858.7)    & 6.05 &  43.37 $\pm$ 0.03 &  320 $\pm$ 20 &  $-$23.11 $\pm$ 0.02  & 5\\
\enddata
\tablecomments{
The Ly$\alpha$ redshifts ($z_{\rm Ly\alpha}$) were determined from the observed wavelengths of the line peak. 
The Ly$\alpha$ luminosities and FWHMs (corrected for instrumental resolution) and the rest-UV continuum magnitudes at 1450 \AA\ ($M_{\rm UV}$) are reported as measured, without correction for dust extinction or IGM absorption.
The last column lists the reference papers; (1) \citet{p1}, (2) \citet{p2}, (3) \citet{p4}, (4) \citet{p10}, (5) \citet{p20}.
}
\end{deluxetable*}


Table \ref{tab:targets} lists the six objects analyzed in this work. 
They were originally identified in the SHELLQs survey through rest-UV spectroscopy, using the Subaru Telescope and the Gran Telescopio Canarias \citep{p1,p2,p4,p10,p20}.
G12 and G13 were discovered and reported in the early phase of the project, and do not quite meet our latest criteria of candidate obscured quasars 
\citep[$L_{\rm Ly\alpha} > 10^{43}$ erg s$^{-1}$ and FWHM $<$ 500 km s$^{-1}$; e.g.,][]{p24}.
We include these two objects in the present analysis as borderline candidates.

The JWST observations were carried out in Cycle 3 as part of the General Observers (GO) program 5645 (the {\em Aether} project; E. Farina et al., 2026, in preparation).
It is a Survey-category program targeting all the 343 quasars at $5.7 < z < 7.0$ that were known at the time of proposal preparation, without any additional selection.\footnote{
A few targets were excluded from the sample due to the lack of suitable guide stars.}
The program used NIRSpec with the high-resolution grating G395H and transmission filter F290LP.
This configuration provides wavelength coverage from 2.87 to 5.14 $\mu$m,
with the spectral resolution $R \sim 2700$.
The observations were carried out in the IFU mode, which provides a 3\arcsec $\times$ 3\arcsec\ square field of view with a spaxel scale of 0\farcs1.
A physical gap between the NIRSpec detectors results in a $\sim$0.1-$\mu$m wavelength gap around 3.98 -- 4.20 $\mu$m, whose exact cutoffs depend on the IFU slice.
The total exposure time on source was 2600 sec per target, divided into 4-point dithers $\times$ 1 integration $\times$ 9 groups.
The CYCLING dither pattern was used with the SMALL spatial scale. 
The readout mode was NRSIRS2.


Following \citet{loiacono24}, we reduced the data using
the JWST NIRSpec official pipeline (CRDS version: 12.1.11, context:
jwst\_1322.pmap). Between Stage 1 (mostly dealing with the ramp
reconstruction) and Stage 2 (dealing with flat fielding and other
instrument-dependent calibrations), we apply an aggressive flagging of
poorly sampled ramps, similar to the TEMPLATES approach
\citep{hutchison24}. Contrary to E. Farina et al. (2026, in preparation) and
R. Decarli et al. (2026, in preparation), after Stage 3 (dealing with the outlier rejection
and the cube reconstruction) we do not remove spectral spikes in an ad hoc manner, 
as the narrow lines of the present targets may be erroneously identified as cosmic rays. The final cubes have
a spatial sampling of $0\farcs1$ per pixel and a spectral sampling of
6.65\,\AA{}\,per channel. The astrometry of the cubes is corrected so
that the quasars lie at their nominal position from the
respective discovery papers.

\section{Results and Discussion} \label{sec:results}

\subsection{Nuclear Emission}

\begin{figure*}
\epsscale{0.9}
\plotone{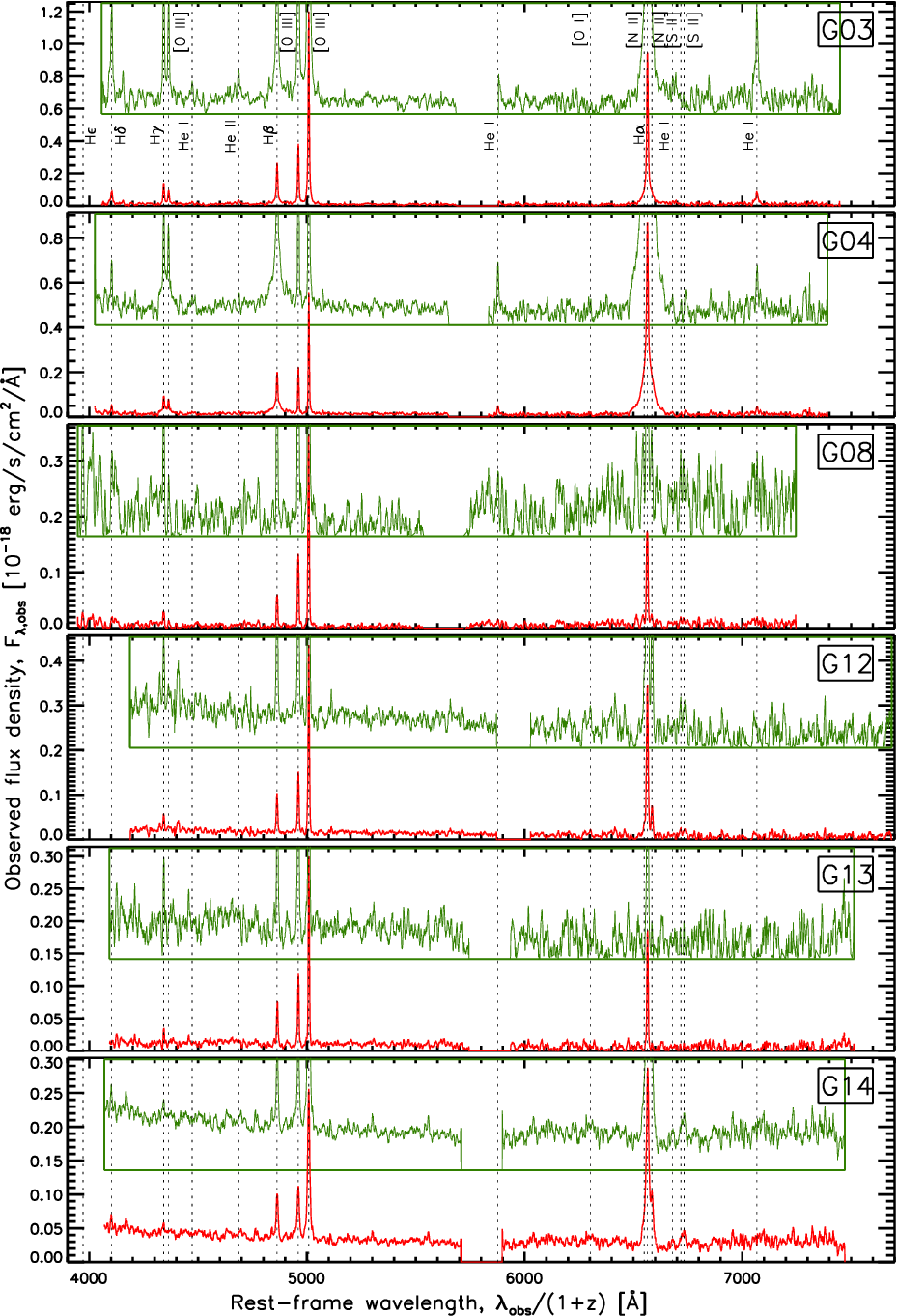}
\caption{
Nuclear spectra extracted from the IFU data cubes for each object (red lines). 
The flux density is given in the observer's frame, while the wavelength is given in the rest frame for the purpose of presentation, 
using the systemic redshifts determined from the spectral models. 
The insets with the green lines show the same spectra plotted on an expanded scale, to make fainter emission features apparent. 
The dotted lines indicate the expected central wavelengths of common emission lines, as labeled in the top panel.
\label{fig:spec}}
\end{figure*}

We extracted the nuclear spectra by integrating the flux over a 7 $\times$ 7 spaxel (0\farcs7 $\times$ 0\farcs7) aperture, centered on the flux-weighted centroid.
We generated Point Spread Function (PSF) models with the JWST Exposure Time Calculator \citep[ETC;][]{jwst_etc},\footnote{
Other {\em Aether} studies typically use the PSF models constructed from the wings of broad emission lines originating from unresolved quasar nuclei
(e.g., R. Decarli et al. 2026, in preparation). We adopt a different approach here, as our objects do not always exhibit broad emission lines.} 
and used them to
correct for chromatic aperture loss \citep{horne86}.
The flux-weighted centroid of each PSF model was matched to that of the observed IFU data cube to sub-pixel accuracy.
The obtained spectra are presented in Figure \ref{fig:spec}.
The formal flux errors provided by the JWST pipeline tend to underestimate the effective noise due to correlated background residuals and other factors. 
We therefore estimated the noise empirically from source-free regions in each wavelength channel, and added it to the formal errors in quadrature.\footnote{
The empirical scatter alone does not fully capture the statistical uncertainty structure propagated through the pipeline processing, particularly in regions where the detector response and source Poisson contribution vary significantly. We therefore conservatively combined the two noise estimates in quadrature, which may overestimate the true noise.
}

\begin{figure*}
\epsscale{1}
\plotone{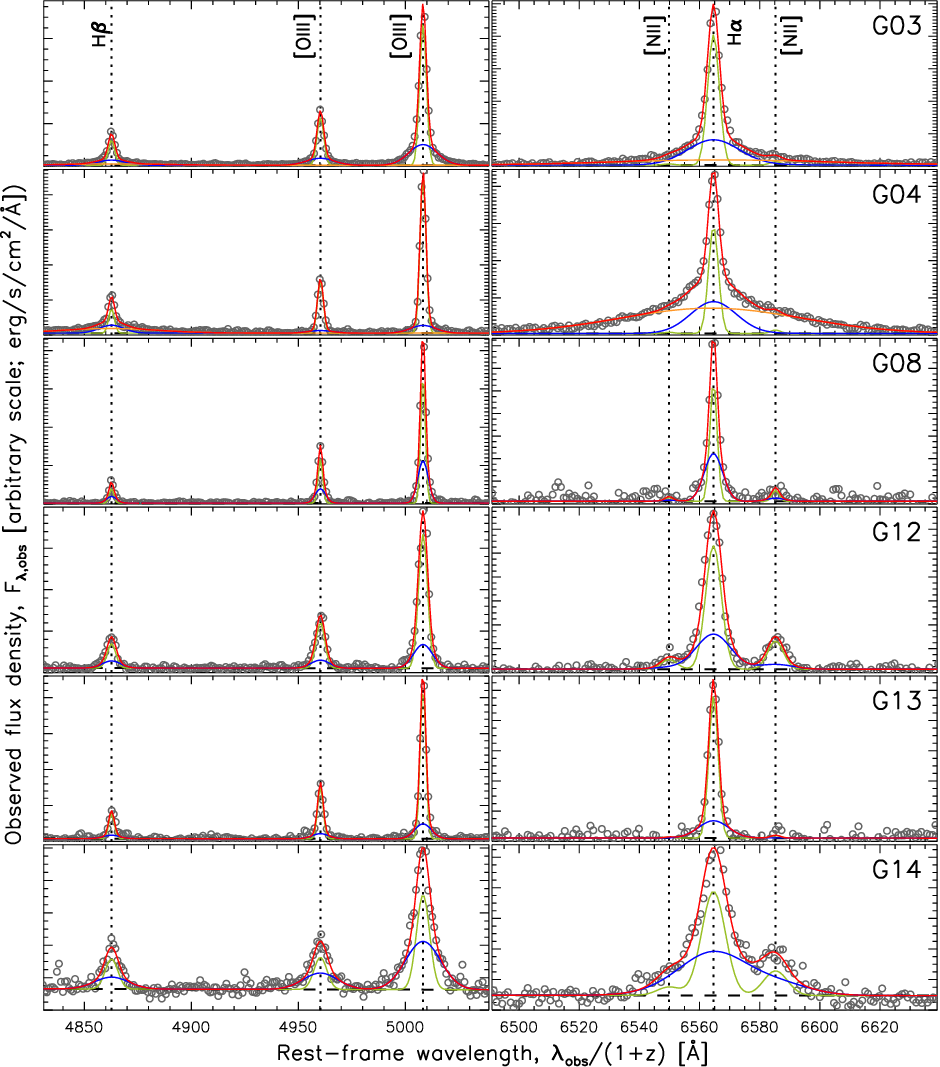}
\caption{The best-fit spectral models (red lines) compared with the observed spectra (open circles)
around H$\beta$ + [\ion{O}{3}] $\lambda$4959, 5007 (left panels) and around H$\alpha$ + [\ion{N}{2}] $\lambda$6585, 6550 (right panels).
The models include a power-law and \ion{Fe}{2} continuum (black dashed lines) and emission lines, each modeled as the sum of two or three Gaussian components (green, blue, and orange lines).
The dotted lines indicate the expected wavelengths of the emission lines, as labeled in the top panels.
\label{fig:specfit}}
\end{figure*}

We modeled and fitted the observed spectra in exactly the same way as in \citet{cy2}.
For details, we refer the reader to that paper; here we briefly summarize the key points.
The continuum emission was assumed to follow a power-law function. 
The \ion{Fe}{2} pseudo-continuum was included in the model, using the \citet{boroson92} template with the velocity broadening tied to emission lines.
This component, however, is very weak in our objects and has little impact on the model fitting.
The forbidden and permitted line profiles were modeled with two and three Gaussian components, respectively.
A third Gaussian component was included only when it has more than 5$\sigma$ significance and
the Bayesian information criterion \citep[BIC; e.g.,][]{liddle07} is reduced by more than 10.
The instrumental line broadening was taken into account, using the disperser resolution as a function of wavelength taken from the JWST User Documentation.\footnote{
https://jwst-docs.stsci.edu}


The model fitting was performed with the least-$\chi^2$ method. 
The best-fit models around H$\beta$ $+$ [\ion{O}{3}] $\lambda$4959, 5007 and around H$\alpha$ are presented in Figure \ref{fig:specfit}.
We found that G08, G12, G13, and G14 have consistent profiles in the forbidden and permitted lines.
On the other hand, 
the presence of a third Gaussian component is strongly favored in the permitted lines of G03 and G04, with a reduction in BIC $\gg$ 10.
Following \citet{cy2}, we decomposed the permitted line fluxes of these two objects into narrow-line (NL) and broad-line (BL) components. 
The NL component is assumed to share the same profile as the forbidden lines, and the remaining emission is attributed to the BL component.
Table \ref{tab:fluxes} in the Appendix \ref{sec:appendix} lists the measured emission line fluxes as well as the properties of other emission components.

We derived the systemic redshift ($z_{\rm sys}$) of each object from the narrowest Gaussian component of emission lines in the best-fit model.
Table \ref{tab:specmodel} reports these values, as well as the FWHMs of the three Gaussian components ($w_1$, $w_2$, and $w_3$) and the rest-frame equivalent widths (EWs) of [\ion{O}{3}] $\lambda$5007, H$\beta$, and H$\alpha$.
The Balmer decrements indicate the dust extinction of $A_V \sim 0 - 2$, 
assuming the intrinsic flux ratio of H$\alpha$/H$\beta$ = 2.86 and the Small Magellanic Cloud extinction law \citep{pei92}.
The table also reports continuum and line luminosities corrected for dust extinction.
We denote the H$\alpha$ luminosities of the NL and BL components as $L_{\rm H\alpha, N}$ and $L_{\rm H\alpha, B}$, respectively.

\begin{figure}
\epsscale{1}
\plotone{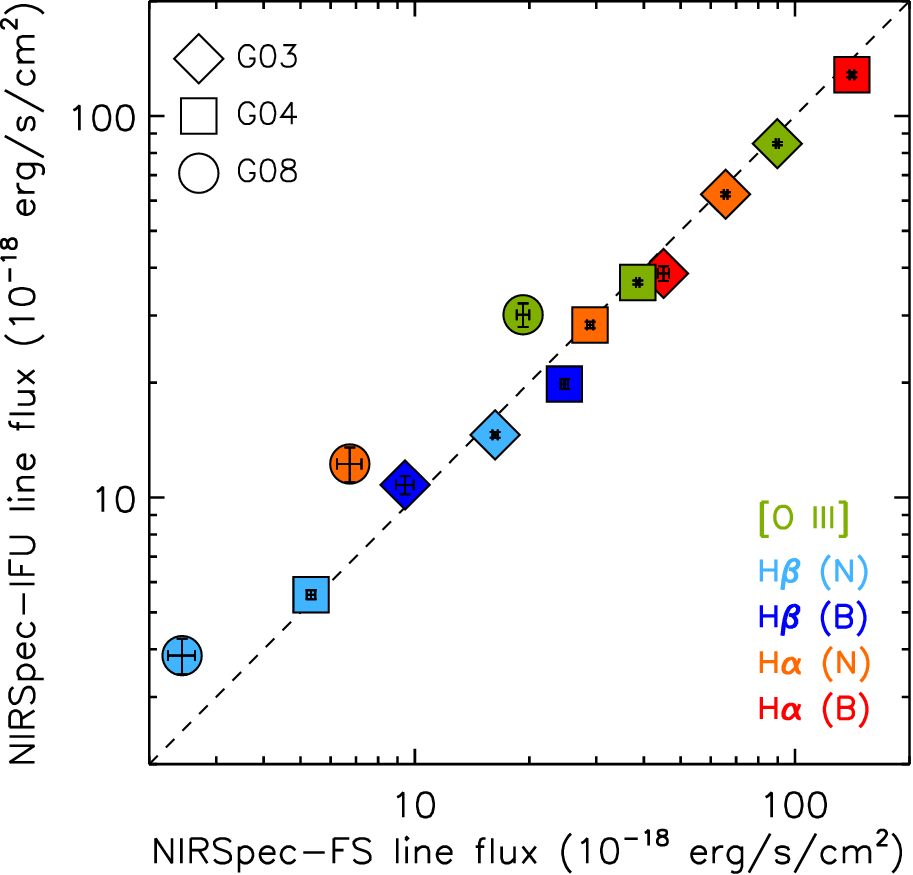}
\caption{
Comparison of the emission line fluxes in the nuclear spectra between the FS and IFU measurements. 
The diamonds, squares, and circles represent G03, G04, and G08, respectively, which were observed in the both modes.
Colors indicate [\ion{O}{3}] $\lambda$5007 (green), the NL/BL components of H$\beta$ (light blue/blue), and the NL/BL components of H$\alpha$ (orange/red).
The dashed line indicates the one-to-one relation.
\label{fig:compare}}
\end{figure}

G03, G04, and G08 were also observed with the NIRSpec Fixed-Slit (FS) mode as part of an independent program GO 3417.
The same grating and transmission filter were used, while the total exposure time on source (3700 sec) was moderately
longer than the present IFU observations.
Figure \ref{fig:compare} compares the [\ion{O}{3}] $\lambda$5007, H$\beta$, and H$\alpha$ line fluxes of the three objects 
between the FS \citep{cy2} and the present IFU measurements.
We observe excellent agreement between the two measurements for G03 and G04, which supports the robustness of the observations, data reduction, and
spectral modeling across different instrumental modes.
On the other hand, the IFU spectrum of G08 yields higher line fluxes than the FS spectrum. 
This is likely because the aperture-loss correction was based on a PSF model, whereas G08 is more spatially extended than G03 and G04 (see \S \ref{sec:extended}).
In what follows, we will adopt the FS measurements for the nuclear spectra of these three objects to take advantage of their higher S/N.

\begin{figure*}
\epsscale{1}
\plotone{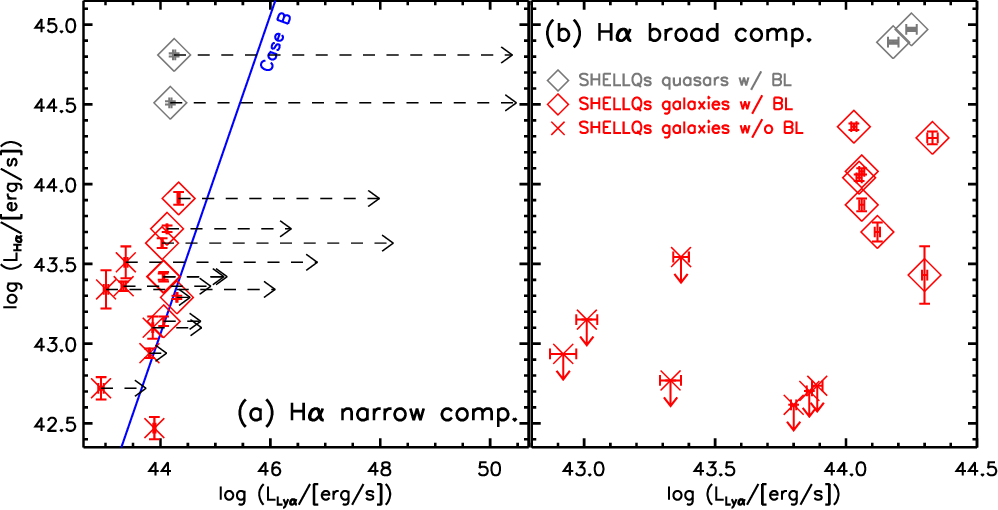}
\caption{Comparison of H$\alpha$ luminosities in the NL (panel a) and BL (panel b) components with Ly$\alpha$ luminosities measured for the SHELLQs objects.
Thirteen objects from \citet{cy2} and three additional objects from the present analysis are shown. 
Galaxies with broad Balmer lines are indicated by red diamonds, those without by red crosses, and two quasars with similar Ly$\alpha$ profiles by gray diamonds.
The downward arrows represent 3$\sigma$ upper limits.
The solid line represents the ratio expected in Case B recombination conditions \citep[e.g.,][]{brocklehurst71, hayes15}.
Note that $L_{\rm H\alpha, N}$ and $L_{\rm H\alpha, B}$ have been corrected for dust extinction, while $L_{\rm Ly\alpha}$ has not.
The dashed arrows in panel a represent the dust-extinction correction for $L_{\rm Ly\alpha}$ inferred from the Balmer decrement.
\label{fig:Lya_vs_Ha}}
\end{figure*}

\begin{deluxetable*}{ccccccc}
\tablecaption{Spectral properties from the best-fit models \label{tab:specmodel}}
\tablewidth{0pt}
\tablehead{
\colhead{} & \colhead{G03} &\colhead{G04} & \colhead{G08} & \colhead{G12} &   \colhead{G13} & \colhead{G14}
}
\startdata
$z_{\rm sys}$           & 6.075                   &   6.130                & 6.272                       &  5.857                &   6.014                 & 6.054  \\
$w_1$ (km s$^{-1}$) & 185 $\pm$    1     &   152 $\pm$    2  &    94  $\pm$   6      &    265 $\pm$    9  &   136 $\pm$    5  &     410 $\pm$   20 \\  
$w_2$ (km s$^{-1}$) &   920 $\pm$ 10    &  810 $\pm$   20  &   320  $\pm$  10    &     610 $\pm$   50 &   530 $\pm$   50 &  1150  $\pm$  50  \\ 
$w_3$ (km s$^{-1}$) &  3850 $\pm$  170 &  3240 $\pm$  50 &  \nodata                  &  \nodata             &  \nodata               &   \nodata \\
EW ([\ion{O}{3}])     & 1090 $\pm$ 100   &   360 $\pm$  30   &   920 $\pm$ 160   &   230 $\pm$  40   &  250 $\pm$  50   &   83 $\pm$   9    \\
EW (H$\alpha$, N) & 970  $\pm$ 70      &   350 $\pm$  20   &   300  $\pm$ 50     & 390 $\pm$  70     &   230 $\pm$  40  & 140 $\pm$  20   \\ 
EW (H$\alpha$, B) &  600 $\pm$  50     &   1570 $\pm$ 100 &   \nodata                  &   \nodata            &   \nodata               &   \nodata   \\
\hline
$A_V (N)$ &   1.09 $\pm$ 0.05 & 1.57 $\pm$ 0.08 & 0.29 $\pm$ 0.29 & 1.28 $\pm$ 0.34  &  0.34 $\pm$ 0.21  & 1.45 $\pm$ 0.29   \\
$A_V (B)$ &   0.62 $\pm$ 0.19 &  2.22 $\pm$ 0.09 &  \nodata &    \nodata &    \nodata &   \nodata \\
\hline
$\log{L_{5100}}$                &  40.77 $\pm$ 0.09 &  41.61 $\pm$ 0.04 &  40.29 $\pm$ 0.13 &  41.17 $\pm$ 0.15 & 40.60 $\pm$ 0.09 & 41.65 $\pm$ 0.13 \\
$\log{L_{\rm [O III]}}$         &  44.04 $\pm$ 0.02 &  43.90 $\pm$ 0.04 & 43.25 $\pm$ 0.13  &  43.55 $\pm$ 0.16 &  43.02 $\pm$ 0.09 &  43.60 $\pm$ 0.14 \\
$\log{L_{\rm H\alpha,N}}$  &  43.75 $\pm$ 0.02 &  43.56 $\pm$ 0.03 &  42.82 $\pm$ 0.10  &  43.34 $\pm$ 0.12 & 42.72 $\pm$ 0.07 &   43.51 $\pm$ 0.10  \\
$\log{L_{\rm H\alpha,B}}$  &  43.39 $\pm$ 0.06  & 44.42 $\pm$ 0.03  & \nodata                  &    \nodata               &   \nodata               &  \nodata\\
\enddata
\tablecomments{
The rest-frame EWs of [\ion{O}{3}] $\lambda$5007 and H$\alpha$ (NL and BL components) are reported in units of \AA.
The luminosities have been corrected for dust extinction and are presented in units of erg s$^{-1}$.
}
\end{deluxetable*}

Figure \ref{fig:Lya_vs_Ha}a compares $L_{\rm H\alpha, N}$ with the Ly$\alpha$ luminosity $L_{\rm Ly\alpha}$.\footnote{
Ly$\alpha$ has a narrow profile overall (FWHM $<$ 500 km s$^{-1}$, as defined by the SHELLQs criteria for candidate obscured quasars),
while there is a hint of a weak broad component in some objects \citep{cy2}.}
Note that $L_{\rm H\alpha, N}$ has been corrected for dust extinction, while $L_{\rm Ly\alpha}$ is as observed, without any correction for dust or IGM absorption.
The figure demonstrates that the observed $L_{\rm Ly\alpha}$/$L_{\rm H\alpha, N}$ ratios are lower than 
expected in Case B recombination conditions \citep[e.g.,][]{brocklehurst71, hayes15}.
On the other hand, using a dust-extinction correction inferred from the Balmer decrement would result in unreasonably high Ly$\alpha$ luminosity in many objects
\citep[$L_{\rm Ly\alpha} = 10^{45 - 50}$ erg s$^{-1}$; e.g.,][]{umeda25}. 
Correcting $L_{\rm Ly\alpha}$ for IGM absorption would make the situation even worse. 
This indicates that the observed Ly$\alpha$ did not suffer the same extinction as H$\alpha$.
In other words, the gas region emitting the hydrogen lines is likely composed of at least two zones.
We speculate that dense gas clouds with moderate dust extinction (nearly opaque to Ly$\alpha$, but allowing partial transmission of H$\alpha$) 
are embedded within the ambient dust-free/poor gas; the observed Ly$\alpha$ originates mostly from the ambient gas, while H$\alpha$ comes from the entire region. 
It is worth noting that Ly$\alpha$ is also subject to resonant scattering, which, together with dust extinction, can significantly modify its spatial distribution. 
This is illustrated by the extended Ly$\alpha$ halos around lower-redshift quasars observed with the Multi Unit Spectroscopic Explorer (MUSE) on the Very Large Telescope (VLT), where emission is found on circumgalactic scales \citep[e.g.,][]{borisova16, ab19}.

\citet{cy2} analyzed the NIRSpec FS spectra of 11 similar galaxies (along with two quasars showing similar Ly$\alpha$ profiles) drawn from the SHELLQs sample, and reported that seven of them exhibit broad rest-optical lines.
We found consistent fitting results for the three common objects with the present work; G03 and G04 have broad lines, while G08 does not.
The additional three objects presented here have somewhat lower $L_{\rm Ly\alpha}$ than these objects, and two of them (G12 and G13) do not show detectable broad Balmer lines.
G14 exhibits broad Balmer lines, but their widths are comparable to those of the forbidden [\ion{O}{3}] lines, suggesting that they trace galaxy-scale gas dynamics
(see Sec \ref{sec:g14} for further discussion).
Figure \ref{fig:Lya_vs_Ha}b summarizes these results; the $L_{\rm H\alpha, B}$ upper limits of the NL objects were estimated by artificially adding a mean BL profile of the BL objects with varying amplitudes and correcting for dust extinction with $A_V (N)$.
Even if the NL objects had an undetectable BL component in H$\alpha$, their luminosities would still be significantly lower than those of the BL objects.
For reference, LRDs and other low-luminosity AGNs discovered by JWST span a broad range of H$\alpha$ luminosities, from $L_{\rm H\alpha,B} \sim 10^{41}$ to $10^{44}$ erg s$^{-1}$ \citep[e.g.,][]{kocevski23, greene24, matthee24}.
Our BL objects occupy the upper end of and partially extend beyond this distribution.

We interpret our BL objects as hosting AGN, whose signature is obscured by dust in the rest-UV spectra.
The FS measurements of the H$\alpha$ BL width and luminosity provide the black hole mass ($M_{\rm BH}$) estimates of $\log (M_{\rm BH}/M_\odot) = 8.21 \pm 0.03$
and $8.46 \pm 0.01$ for G03 and G04, respectively \citep{cy2}.\footnote{
E. Farina et al. (2026, in preparation) will provide consistent $M_{\rm BH}$ measurements across the whole {\em Aether} sample, including these two objects.}
On the other hand, \citet{iwasawa25} reported that G03 and G04, along with two similar objects (G01 and G02), were not detected in their Chandra observations, which indicate significantly lower X-ray to rest-optical luminosity ratios than those of normal AGNs.
This is similar to the situation observed in LRDs, as described in \S \ref{sec:intro}.

We have further examined radio images from the LOFAR Two-metre Sky Survey \citep[LoTSS;][]{shimwell17} at 144 MHz 
and the Very Large Array Sky Survey \citep[VLASS;][]{lacy20} at 3 GHz, but none of the present six objects (only four of them are found in the VLASS footprint) is individually detected. 
They also remain undetected in the stacked data, which reach the $3\sigma$ luminosity limits of $7.3 \times 10^{25}$ W Hz$^{-1}$ (LoTSS) and $5.0 \times 10^{25}$ W Hz$^{-1}$ (VLASS). 
The LoTSS stack provides a $3\sigma$ upper limit on the radio loudness, $L_\nu$ (1.4GHz)/$L_\nu$(2500\AA) $<$ 1300, assuming a radio spectral index of $\alpha = 0.7$. 
This limit is significantly higher than the conventional threshold between radio-loud and radio-quiet AGNs. 
Much deeper observations are required to constrain their nature based on the radio emission.

As the NIRSpec-observed sample grows, it is becoming clear that the presence of an AGN is closely linked to Ly$\alpha$ luminosity;
broad rest-optical lines appear ubiquitous at $L_{\rm Ly\alpha} > 10^{44}$ erg s $^{-1}$ and rapidly disappear toward lower luminosities.
At face value, this indicates AGN fractions of $>$77 \% (7 out of 7 objects; 1$\sigma$ confidence level under binomial statistics) and $<$15 \% (0 out of 7 objects) above and below $L_{\rm Ly\alpha} = 10^{44}$ erg s$^{-1}$, respectively.
The fairly sharp transition in the AGN fraction is intriguing.
It indicates that the LF of non-AGN Ly$\alpha$ emitters (LAEs) intersects the AGN LF at $L_{\rm Ly\alpha} \sim 10^{44}$ erg s $^{-1}$,
above which AGNs become the dominant population.
This transition luminosity corresponds to a star formation rate (SFR) of $\sim$100 $f_{\rm esc}^{-1}$ $M_\odot$ yr$^{-1}$, 
assuming Case B recombination conditions and the H$\alpha$-SFR relation of \citet{kennicutt98}.
Here $f_{\rm esc}$ represents the Ly$\alpha$ photon escape fraction, whose typical value is 0.05 -- 0.2 at $z \sim 6$ \citep[e.g.,][]{tang23,chen24,saxena24}.

At lower redshifts ($z \sim 2$), it has been reported that the AGN fraction among LAEs increases more gradually,
from nearly 0 \% at $L_{\rm Ly\alpha} = 10^{42.5}$ erg s$^{-1}$ to 100 \% at $10^{43.5}$ erg s$^{-1}$ \citep{sobral18, spinoso20}.
The higher LAE number density at $z \sim 6$ \citep[e.g.,][]{konno16} may be responsible for the higher transition luminosity, where the steep fall-off of the (non-AGN) LAE LF 
may cause the sharp transition.
In addition, the IGM \ion{H}{1} opacity is much higher at $z \sim 6$ than at $z \sim 2$. 
Hard radiation fields can create large ionized bubbles preferentially around AGNs, which may help Ly$\alpha$ photons escape IGM absorption.
Alternatively, dust extinction may be more significant at the lower (observed) Ly$\alpha$ luminosities, obscuring AGN signatures at both rest-UV and rest-optical wavelengths.
Indeed, Figure \ref{fig:Lya_vs_Ha}a shows that the $L_{\rm Ly\alpha}/L_{\rm H\alpha}$ ratios of the objects without broad Balmer lines are further offset from the Case B value than those with broad Balmer lines.

\begin{figure}
\epsscale{1}
\plotone{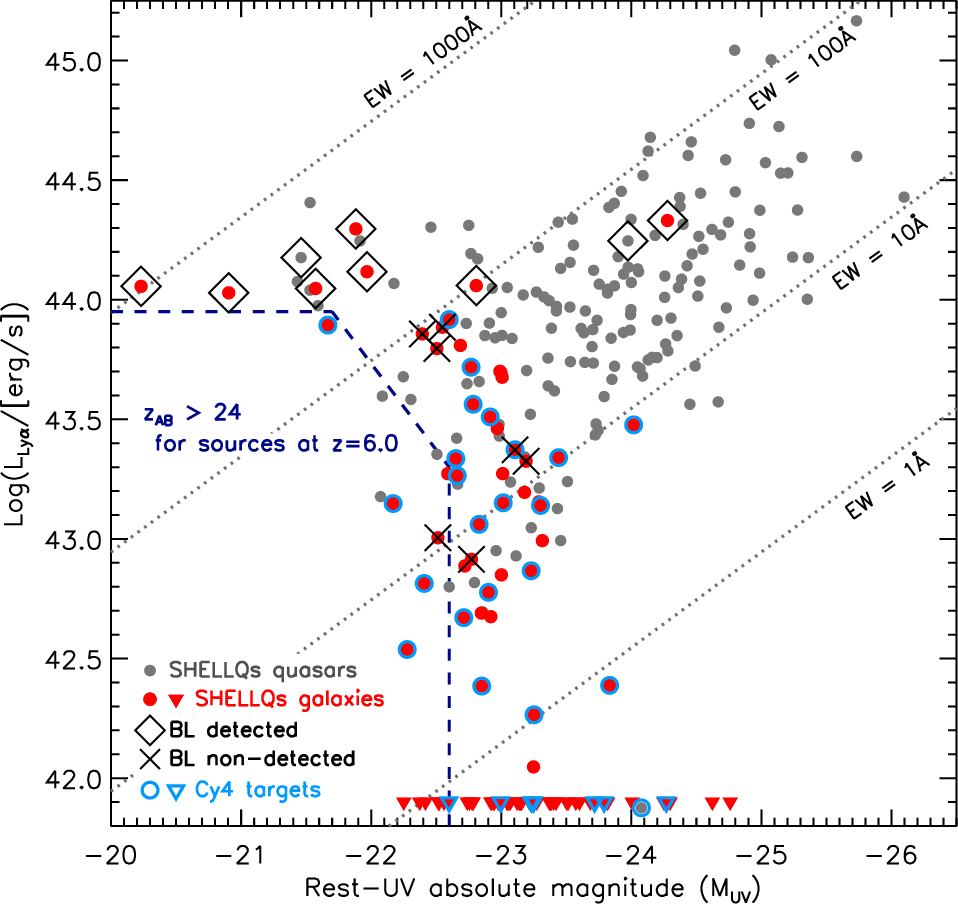}
\caption{Rest-UV continuum magnitudes and Ly$\alpha$ luminosities of all SHELLQs quasars (gray dots) and galaxies (red dots; those without Ly$\alpha$ detection
are represented by red triangles, plotted at an arbitrary vertical position). 
 The dotted lines represent constant rest-frame EWs of 1, 10, 100, and 1000 \AA. 
Sources to the left and bottom of the dashed lines are fainter than the SHELLQs limiting magnitude ($z_{\rm AB}$ = 24.0) at z = 6.0.
The objects with existing JWST/NIRSpec measurements are marked by diamonds (those with broad lines) and crosses 
(those without detectable broad lines).
The GO 7491 targets currently being observed in Cycle 4 are outlined in blue; see \S \ref{sec:summary} in the text.
\label{fig:Muv_vs_Lya}}
\end{figure}

Nevertheless, we note that the above arguments are subject to several selection biases. 
Since the primary goal of SHELLQs was to discover unobscured quasars with strong, unresolved continuum emission, its photometric selection was not 
optimized for obscured quasars, 
either in terms of color or spatial extent.
Figure \ref{fig:Muv_vs_Lya} demonstrates that the selection becomes incomplete particularly at $L_{\rm Ly\alpha} < 10^{44}$ erg s$^{-1}$, due to the adopted magnitude cut of $z_{\rm AB} < 24$.
At a given $L_{\rm Ly\alpha}$, objects with higher Ly$\alpha$ EWs tend to be excluded from the selection, because of their fainter continuum (and thus total) emission.
We also note that the Ly$\alpha$ luminosities have not been corrected for unknown dust extinction and/or IGM absorption; such a correction could yield significantly higher
$L_{\rm Ly\alpha}$ than presented here.
Finally, the AGN fraction considered here includes only objects with detectable broad Balmer lines. 
AGNs with heavier dust obscuration are not included and may therefore be missed. 
Such a population could be identified using other AGN tracers, such as high-ionization lines or multi-wavelength properties; this remains for future studies (e.g., K. Aoki et al. 2026, in preparation).

\subsection{Extended Emission} \label{sec:extended}

\begin{figure*}
\epsscale{0.8}
\plotone{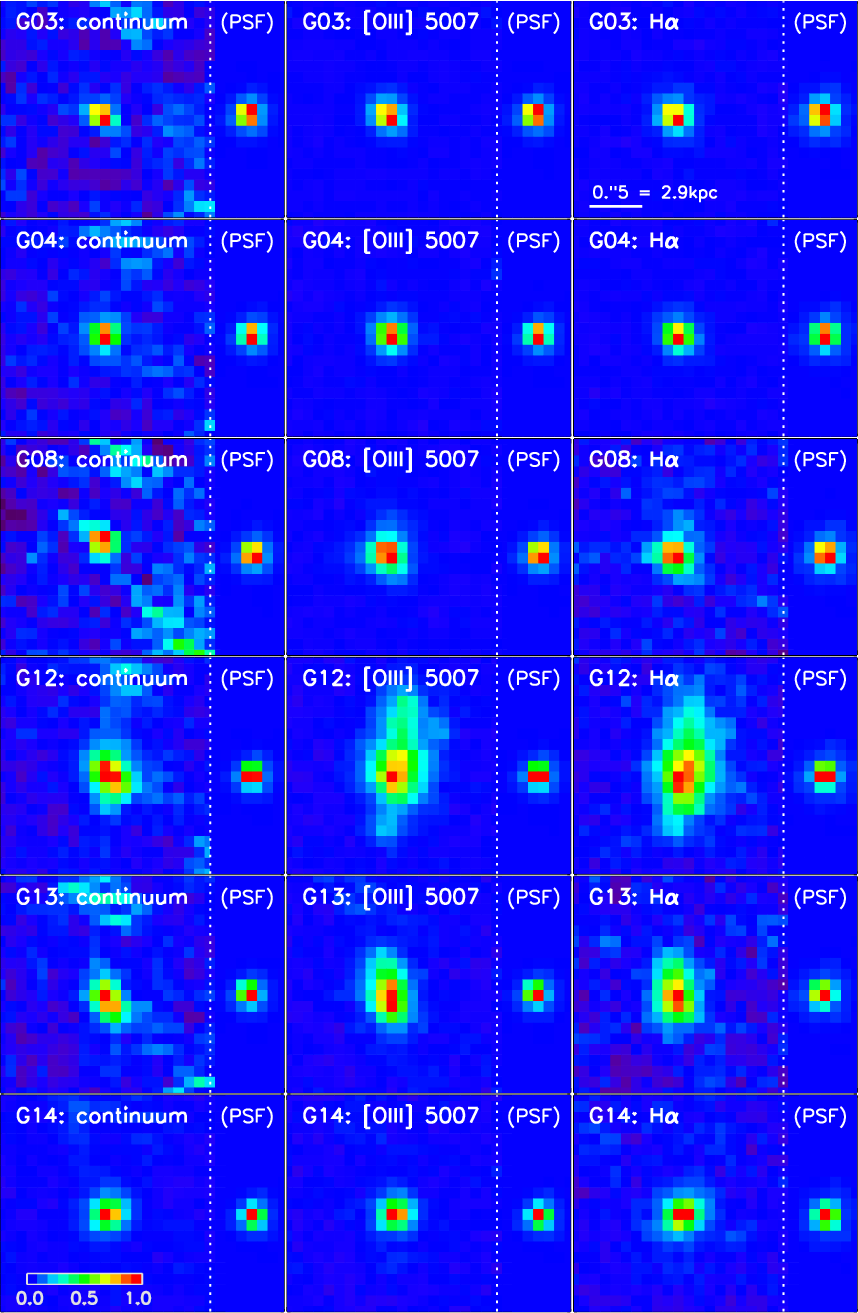}
\caption{NIRSpec IFU maps for the six galaxies. 
Colors represent peak-normalized flux on a linear scale, increasing from blue to red (color bar is shown in the bottom left panel).
Each row corresponds to one galaxy (G03, G04, G08, G12, G13, and G14 from top to bottom), 
and the panels from left to right show the continuum emission integrated over $\lambda_{\rm rest} = 4200 - 5500$ \AA, 
the [\ion{O}{3}] $\lambda$5007 line, and the H$\alpha$ line, respectively.
In each panel, the corresponding PSF model is shown to the right of the object image.
Each panel covers 2\arcsec $\times$ 2\arcsec, compared to the full IFU field of view of 3\arcsec $\times$ 3\arcsec.
The scale bar in the top right panel represents 0\farcs5 (2.9 kpc in proper distance at $z = 6$).
North is up and East is to the left.
\label{fig:map_atlas1}}
\end{figure*}

We now turn our attention to spatially extended emission in the IFU data cube.
Figure \ref{fig:map_atlas1} presents the continuum, [\ion{O}{3}] $\lambda$5007, and H$\alpha$ emission maps centered on each of the six objects.
The continuum emission was integrated over $\lambda_{\rm rest} = 4200 - 5500$ \AA, after masking emission lines.
We performed 5$\sigma$ clipping at each wavelength channel, and then used the inverse-variance-weighted mean for flux integration at each spaxel. 
Apparent snowball features were identified and removed by visual inspection.
Finally, detector-based stripe patterns were removed following the recipe in \citet{decarli24}.
The line emission maps were created in a similar manner, but without 5$\sigma$ clipping.
This is because the [\ion{O}{3}] and H$\alpha$ lines are sufficiently strong to be detected in some individual wavelength channels.
While some cosmic ray impacts may leak through this process, their effect is expected to be minimal, as the emission lines span a much narrower velocity range than the continuum, 
and obvious snowball features were removed manually.
The line fluxes were simply summed over the velocity range corresponding to the FWHM of the second Gaussian component in the best-fit model 
(i.e., $w_2$ in Table \ref{tab:specmodel}).


\begin{deluxetable}{crrr}
\tablecaption{Source concentration relative to PSF \label{tab:trace}}
\tablewidth{0pt}
\tablehead{
\colhead{Object} & \colhead{Continuum} & \colhead{[\ion{O}{3}]} & \colhead{H$\alpha$}
}
\startdata
G03    &  1.04 $\pm$  0.04 &  0.94 $\pm$  0.01 &  0.96 $\pm$  0.01 \\
G04    &  0.88 $\pm$  0.02 &  0.92 $\pm$  0.01 &  0.97 $\pm$  0.01 \\
G08    &  0.97 $\pm$  0.04 &  0.83 $\pm$  0.01 &  0.84 $\pm$  0.02 \\
G12    &  0.69 $\pm$  0.01 &  0.59 $\pm$  0.01 &  0.58 $\pm$  0.01 \\
G13    &  0.78 $\pm$  0.01 &  0.70 $\pm$  0.01 &  0.68 $\pm$  0.01 \\
G14    &  0.86 $\pm$  0.01 &  0.88 $\pm$  0.01 &  0.81 $\pm$  0.01 \\
\enddata
\tablecomments{
This table lists the source concentration quantified by $R_{35} / R_{35}^{\rm PSF}$ (see text)
for continuum at $\lambda_{\rm rest} = 4400 - 5200$ \AA, [\ion{O}{3}] $\lambda$5007, and H$\alpha$.
}
\end{deluxetable}

Figure \ref{fig:map_atlas1} demonstrates a wide range of apparent morphologies, from unresolved to significantly extended. 
This is remarkable, given that all the objects were identified as point sources at the resolution of Subaru/HSC ($\sim$0\farcs7 at rest-UV wavelengths).
We measured the flux ratios between 3 $\times$ 3 and 5 $\times$ 5 spaxel apertures ($R_{35}$) as a rough indicator of source concentration, and
compared them with those of the PSF model generated with the JWST ETC  ($R_{35}^{\rm PSF}$). 
Table \ref{tab:trace} lists the ratios $R_{35} / R_{35}^{\rm PSF}$.
The uncertainties were estimated by repeating the measurements after adding random flux fluctuations to the images based on the associated error maps.
The table shows that G03 and G04 exhibit compact morphology overall, which are consistent with the PSF. 
Together with the presence of broad emission lines, these measurements indicate that both the continuum and line emission are dominated by unresolved AGN.
In contrast, the remaining four objects without broad lines tend to be more extended.
In particular, the significant extents of the line emission indicate both a substantial reservoir of interstellar gas and an extended ionizing radiation field. 
This is most significant in G12 and G13, in which the line emission reaches projected extents of up to 10 kpc and 5 kpc (in proper distance), respectively.
Assuming that the H$\alpha$ photons are produced in \ion{H}{2} regions around massive stars, the inferred SFRs are 
$60 \pm 10$, $570 \pm 140$, $70 \pm 10$, and $280 \pm 60$ $M_\odot$ yr$^{-1}$ 
in G08, G12, G13, and G14, respectively, using the \citet{kennicutt98} relation. 
The SFRs have been corrected for dust extinction, assuming the $A_V (N)$ values reported in Table \ref{tab:specmodel}.
The continuum emission tends to be less extended, which may indicate the concentration of stellar populations within the central few pixels ($\sim$1 kpc across) and/or 
the presence of some AGN radiation from the unresolved nuclei. 


\begin{figure*}
\epsscale{0.6}
\plotone{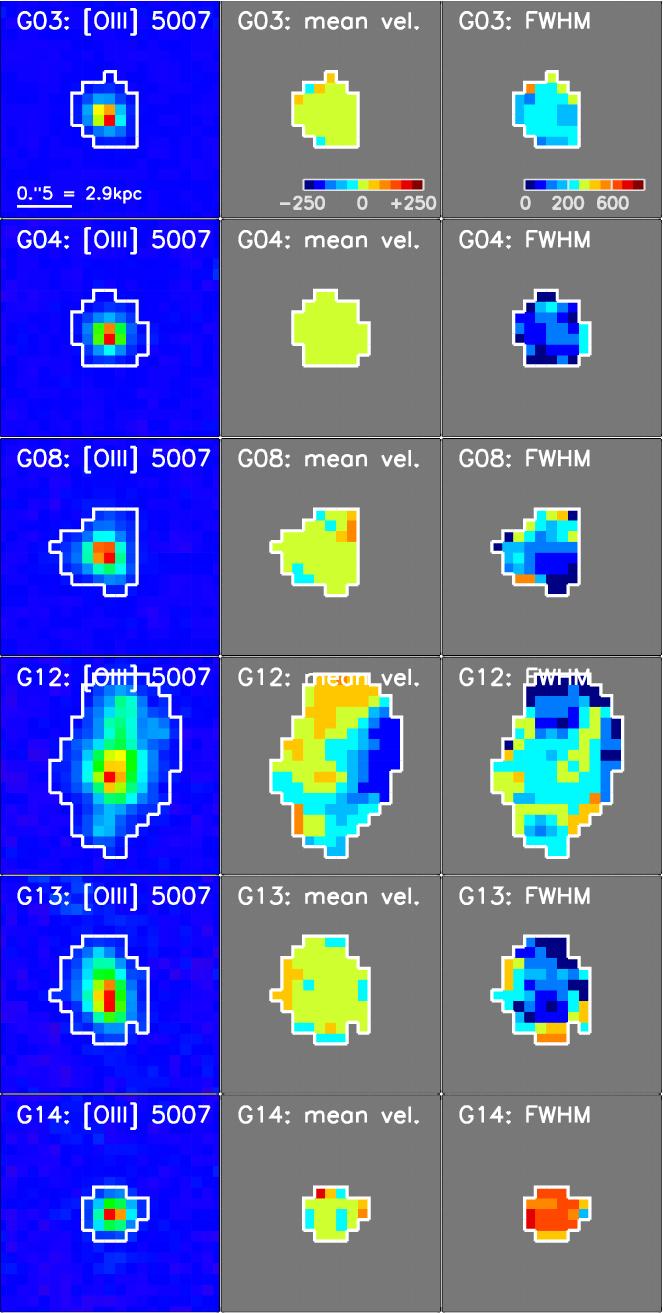}
\caption{NIRSpec IFU maps for the six galaxies. Each row corresponds to one galaxy (G03, G04, G08, G12, G13, and G14 from top to bottom), 
and the panels from left to right show the maps of integrated flux (equivalent to the middle-column panels of Figure \ref{fig:map_atlas1}, including some continuum contribution), mean velocity, and velocity width (FWHM) of the [\ion{O}{3}] $\lambda$5007 emission. 
Spaxels used for the velocity measurements are outlined by the thick white lines. 
In the outer regions of the galaxies, we binned several (up to five) spaxels to increase S/N for the velocity measurements.
The scale bar in the top left panel corresponds to 0\farcs5 (2.9 kpc in proper distance at $z = 6$).
In the middle-column panels, the color scale represents 11 velocity bins ranging from $-250$ to $+250$ km s$^{-1}$, with a spacing of 50 km s$^{-1}$.
In the right panels, the bluest four colors correspond to 50 km s$^{-1}$ velocity bins spanning 0 -- 200 km s$^{-1}$, 
the middle four to 100 km s$^{-1}$ bins spanning 200 -- 600 km s$^{-1}$, 
and the reddest three to 200 km s$^{-1}$ bins spanning 600 -- 1200 km s$^{-1}$.
The corresponding color bars are shown in the top middle and top right panels.
\label{fig:map_atlas2}}
\end{figure*}

We measured the kinematics of the ionized gas using [\ion{O}{3}] $\lambda$5007 emission, and present the maps of mean velocity and velocity width
(FWHM, corrected for instrumental broadening) in Figure \ref{fig:map_atlas2}.
The [\ion{O}{3}] line was chosen because it is brighter than H$\alpha$ (the NL component for G03 and G04) and the background continuum emission is more spatially uniform at shorter wavelengths.
G04, G08, and G13 exhibit similar gas kinematics, with little spatial structure of the mean velocity and moderate velocity width up to $\sim$200 km s$^{-1}$.
G03 has a similar structure, but with slightly elevated velocity width of 200 -- 300 km s$^{-1}$. 
A signature of galaxy rotation may be visible only in G12, where the northern and western regions exhibit significant redshift (up to 100 km s$^{-1}$) and blueshift (up to 200 km s$^{-1}$), respectively, relative to the systemic velocity.
This pattern is somewhat inconsistent with the naive expectation that, if the apparent elongation of the galaxy is due to inclination, the velocity gradient would align with the major axis oriented north-south.
This discrepancy may be partly due to star formation driven outflows, which could perturb the global velocity field from simple rotational motion.
Finally, G14 exhibits peculiar kinematics, characterized by extremely high velocity width reaching almost 1000 km s$^{-1}$.
We discuss this object further in the next section.

\subsection{$J$121503.42$-$014858.7 (G14)} \label{sec:g14}
 
 G14 was previously reported by \citet{p20} to be part of a physical quasar pair. 
 The pair was identified as consisting of a brighter, more compact object (``C1" with $M_{1450} = -23.11$, corresponding to G14 in this paper) and 
 a fainter, more extended companion (``C2" with $M_{1450} = -22.66$).
The two objects exhibit strong Ly$\alpha$ lines at almost identical wavelengths, which provide a common redshift of $z = 6.05$.
Their angular separation (2\farcs0) corresponds to a projected proper distance of 12 kpc, suggesting that they are physically associated.
Indeed, a bridging gas structure was detected in Ly$\alpha$ emission.
Follow-up ALMA observations \citep{izumi24} further revealed spectacular [\ion{C}{2}] 158-$\mu$m emission bridging and surrounding the two objects, providing clear evidence for their dynamical interaction.

The two objects each show a relatively broad Ly$\alpha$ component with FWHM of 1500 km s$^{-1}$ (G14) and 1300 km s$^{-1}$ (C2).
Their continuum and Ly$\alpha$ luminosities are much higher than those of typical galaxies, making it unlikely that such luminous objects would form a close pair by chance.
Based on these properties, \citet{p20} proposed that both objects are likely quasars. 
However, the rest-optical spectrum of G14 (see Figures \ref{fig:spec} and \ref{fig:specfit}) reveals that its [\ion{O}{3}] lines are similarly broad.
Both permitted and forbidden lines are well fitted with a combination of two Gaussians, and the FWHM of the broader component ($w_2 = 1200$ km s$^{-1}$; see Table \ref{tab:specmodel}) is consistent with that of the broad Ly$\alpha$ component.
The [\ion{C}{2}] line from the ALMA observations also exhibits a broad component with a similar width, FWHM = 1200 km s$^{-1}$.
Therefore, we conclude that the broad line profile of G14 is not produced by gas motion due to the gravity of an SMBH, and thus it is unclear whether this object hosts a quasar. 
Its [\ion{O}{3}] $\lambda$5007/H$\beta$ ratio ($\sim$3.7) is moderately high, but this alone cannot securely classify the object as a quasar, as JWST has revealed 
high-$z$ star-forming galaxies with similarly strong high ionization lines \citep[e.g.,][]{scholtz25}.

Figure \ref{fig:map_atlas2} shows that G14 has extremely high velocity width compared to the other galaxies.
Normal star-forming galaxies are not expected to produce line profiles significantly broader than 500 km s$^{-1}$, even in the presence of galactic outflows
\citep[e.g.,][]{newman12,swinbank19}.
The turbulent gas kinematics of G14 may be attributed to dynamical interaction with the companion galaxy C2, located to the east-southeast.
The present NIRSpec data cube covers C2 at the edge of the IFU field of view. 
A detailed analysis of the extended emission in the G14-C2 system will be presented in S. Onorato et al. (2026, in preparation).

\section{Summary and Conclusions} \label{sec:summary}

This paper presents the NIRSpec IFU observations of six galaxies at $z \sim 6$, obtained as part of the JWST GO 5645 program (the {\em Aether} project).
The targets were originally identified by the SHELLQs project, as candidate obscured quasars
with luminous ($\gtrsim 10^{43}$ erg s$^{-1}$) but narrow ($\lesssim$ 500 km s$^{-1}$) Ly$\alpha$ emission.
The NIRSpec spectra revealed that two objects (G03 and G04) have broad line components (FWHM $> 3000$ km s$^{-1}$) 
only in permitted Balmer lines, indicating the presence of AGN.
The remaining four objects (G08, G12, G13, and G14) show similar profiles in the permitted and forbidden lines.

We combined the above sample with similar SHELLQs objects presented in \citet{cy2}, and analyzed their Ly$\alpha$ and H$\alpha$ luminosities.
Overall, we found that the dust-extinction correction inferred from the Balmer decrement would result in unreasonably high values of $L_{\rm Ly\alpha}$.
This may indicate that the line-emitting gas consists of multiple zones.
We also found that the presence of broad permitted lines is strongly coupled with Ly$\alpha$ luminosity.
The inferred AGN fraction is $>$77 \% and $<$15 \% (1$\sigma$ confidence level) above and below $L_{\rm Ly\alpha} = 10^{44}$ erg s$^{-1}$, respectively.
The fairly sharp transition in AGN fraction may be explained by the relatively high LAE number density at $z \sim 6$ and a sharp fall-off of the (non-AGN) LAE LF at the luminous end.
The IGM and/or dust absorption of Ly$\alpha$ photons may also be at work.

The extended emission of the six objects exhibits a wide range of apparent morphologies. 
The two AGN hosts (G03 and G04) have compact morphology, while the other four objects exhibit more extended emission.
In particular, the ionized gas of G12 and G13 extends to 10 kpc and 5 kpc in diameter, respectively.
The total SFRs estimated from the H$\alpha$ luminosities of the four objects range from 60 to 600 $M_\odot$ yr$^{-1}$.
G03, G04, G08, and G13 show little ordered kinematics, with moderate velocity widths (FWHM) up to 200 -- 300 km s$^{-1}$.
A signature of rotation may be visible only in G12. 
Finally, G14 exhibits an extremely high velocity width, reaching 1000 km s$^{-1}$. 
While this object was thought to be a quasar, 
such a large width of the forbidden [\ion{O}{3}] line calls into question whether there is a broad line region in this object.

Combined with the previous work by \citet{cy2}, the present analysis reveals the diverse nature of candidate obscured quasars discovered by SHELLQs.
Broad permitted lines, indicative of AGN activity, appear almost ubiquitously at the highest Ly$\alpha$ luminosities.
Objects with weaker Ly$\alpha$ are more likely dominated by star-forming galaxies, some of which exhibit considerably extended regions of ionized gas and 
total SFR well above 100 $M_\odot$ yr$^{-1}$.
The possible presence of obscured AGNs in these objects, which may be traced by high ionization lines or multi-wavelength properties, is left for future studies (e.g., K. Aoki et al. 2026, in preparation).
Those candidate obscured quasars occupy the intermediate luminosity regime between classical quasars and the low-luminosity AGNs discovered by JWST, including LRDs, potentially bridging the two populations.
JWST/NIRSpec observations of 30 additional SHELLQs objects (see Figure~\ref{fig:Muv_vs_Lya}) are planned in the Cycle 4 GO 7491 program, promising a more complete exploration of the diversity within this population.

\acknowledgments


This work is based on observations made with the NASA/ESA/CSA James Webb Space Telescope. 
The data were obtained from the Mikulski Archive for Space Telescopes (MAST) at the Space Telescope Science Institute (STScI), which is operated by the Association of Universities for Research in Astronomy, Inc., under NASA contract NAS 5-03127 for JWST. 
These observations are associated with the program GO 5645.

The GO 5645 data of the six objects described here may be obtained from the MAST archive at
\dataset[doi:10.17909/yvyw-8667]{https://dx.doi.org/10.17909/yvyw-8667}.
The data available there, however, were pipeline-processed at the STScI, and are therefore not identical to those presented in this paper.




Y. M. was supported by the Japan Society for the Promotion of Science (JSPS) KAKENHI Grants No. 21H04494 and No. 26K00745. 
C. M. acknowledges support from Fondecyt Iniciacion grant 11240336 and the ANID BASAL project FB210003.

\clearpage

\appendix

\section{Spectral measurements} \label{sec:appendix}

The emission line fluxes and continuum properties of the six objects were measured from the best-fit spectral models, and are reported in the observer's frame in Table \ref{tab:fluxes}.

\begin{deluxetable*}{ccccccc}
\tablecaption{Emission line fluxes and continuum properties \label{tab:fluxes}}
\tablewidth{0pt}
\tablehead{
\colhead{} & \colhead{G03} &\colhead{G04} & \colhead{G08} & \colhead{G12} &   \colhead{G13} & \colhead{G14} \\
}
\startdata
\hline\multicolumn{7}{c}{Emission lines, NL component}\\\hline
                                    H6 &  3.47 $\pm$  0.25 &   1.10 $\pm$  0.14 &   0.78 $\pm$  0.16 &   \nodata                &   \nodata                &   1.72 $\pm$  0.35 \\
                                    H5 &  6.57 $\pm$  0.21 &   2.24 $\pm$  0.12 &   1.69 $\pm$  0.22 &   2.00 $\pm$  0.30 &   1.11 $\pm$  0.15 &   1.32 $\pm$  0.27 \\
                              H$\beta$ & 14.59 $\pm$  0.25 &   5.56 $\pm$  0.15 &   3.85 $\pm$  0.41 &   5.10 $\pm$  0.62 &   3.19 $\pm$  0.25 &   5.71 $\pm$  0.62 \\
                             H$\alpha$ & 62.34 $\pm$  0.78 &  28.33 $\pm$  0.56 &  12.24 $\pm$  1.28 &  23.35 $\pm$  2.76 &  10.32 $\pm$  0.72 &  27.83 $\pm$  2.87 \\
             \ion{He}{1} $\lambda$4473 &  0.69 $\pm$  0.14 &   $<$0.43               &   $<$0.24               &   $<$0.49               &   $<$0.22               &   $<$0.61 \\
             \ion{He}{2} $\lambda$4687 &  1.04 $\pm$  0.16 &   0.28 $\pm$  0.08 &   $<$0.38               &   $<$0.50               &   $<$0.50               &   $<$1.12 \\
             \ion{He}{1} $\lambda$5877 &  7.12 $\pm$  1.52 &   1.31 $\pm$  0.11 &   0.63 $\pm$  0.17 &   \nodata                &   \nodata                 &   \nodata  \\
             \ion{He}{1} $\lambda$6679 &  $<$1.40               &   $<$0.82               &   $<$0.76              &   $<$0.74               &   $<$0.47                &   $<$1.45 \\
             \ion{He}{1} $\lambda$7067 &  4.35 $\pm$  0.33 &   1.00 $\pm$  0.19 &   1.02 $\pm$  0.27 &   $<$0.93               &   $<$0.24                 &   2.05 $\pm$  0.54 \\
        $[$\ion{O}{3}$]$ $\lambda$4363 &  4.93 $\pm$  0.20 &   2.81 $\pm$  0.15 &   0.51 $\pm$  0.12 &   $<$1.06                &   $<$0.28               &   $<$1.14  \\
        $[$\ion{O}{3}$]$ $\lambda$5007 & 84.62 $\pm$  0.83 &  36.61 $\pm$  0.45 &  30.13 $\pm$  2.15 &  24.72 $\pm$  2.26 &  18.35 $\pm$  0.78 &  21.28 $\pm$  2.27 \\
        $[$\ion{O}{1}$]$ $\lambda$6300 &  $<$0.16               &   $<$0.75               &   $<$0.57              &   0.72 $\pm$  0.19 &   $<$0.46                &   1.26 $\pm$  0.35 \\
        $[$\ion{N}{2}$]$ $\lambda$6585 &  1.67 $\pm$  0.30 &   0.91 $\pm$  0.23 &   0.95 $\pm$  0.19 &   4.50 $\pm$  0.44 &   $<$0.69                &   7.56 $\pm$  0.87 \\
        $[$\ion{S}{2}$]$ $\lambda$6716 &  $<$0.27               &   $<$0.92                &   0.67 $\pm$  0.21 &   0.95 $\pm$  0.24 &   $<$0.24              &   $<$1.93 \\
        $[$\ion{S}{2}$]$ $\lambda$6731 &  $<$0.37               &   $<$0.48                &   0.61 $\pm$  0.20 &   0.85 $\pm$  0.24 &   $<$0.74               &   2.12 $\pm$  0.46 \\
\hline\multicolumn{7}{c}{Emission lines, BL component}\\\hline
                                    H6 &  4.33 $\pm$  0.85 &   $<$3.25               &   \nodata                &   \nodata                &   \nodata                &   \nodata  \\
                                    H5 &  2.47 $\pm$  0.65 &   6.63 $\pm$  0.60 &   \nodata                &   \nodata                &   \nodata                &   \nodata  \\
                              H$\beta$ & 10.79 $\pm$  0.60 &  19.83 $\pm$  0.58 &   \nodata                &   \nodata                &   \nodata                &   \nodata  \\
                             H$\alpha$ & 38.67 $\pm$  1.71 & 128.22 $\pm$  1.94 &   \nodata                &   \nodata                &   \nodata                &   \nodata  \\
             \ion{He}{1} $\lambda$4473 &  $<$0.99               &   $<$0.55               &   \nodata                &   \nodata                &   \nodata                &   \nodata  \\
             \ion{He}{2} $\lambda$4687 &  2.91 $\pm$  0.55 &   $<$0.99               &   \nodata                &   \nodata                &   \nodata                &   \nodata  \\
             \ion{He}{1} $\lambda$5877 &  1.13 $\pm$  0.27 &   1.26 $\pm$  0.16 &   \nodata                &   \nodata                &   \nodata                &   \nodata  \\
             \ion{He}{1} $\lambda$6679 &  5.60 $\pm$  0.92 &   $<$1.31               &   \nodata                &   \nodata                &   \nodata                &   \nodata  \\
             \ion{He}{1} $\lambda$7067 &  8.69 $\pm$  1.14 &   5.20 $\pm$  1.02 &   \nodata                &   \nodata                &   \nodata                &   \nodata  \\
        \ion{Fe}{2} $\lambda$4434-4684 &  $<$3.21           &   $<$2.58               &   \nodata                &   \nodata                &   \nodata                &   \nodata  \\
\hline\multicolumn{7}{c}{Power-law continuum}\\\hline
              $F_{5100}$  & 10.86 $\pm$  0.29 &  14.05 $\pm$  0.26 &   4.57 $\pm$  0.14 &  15.33 $\pm$  0.15 &  10.09 $\pm$  0.14 &  35.50 $\pm$  0.15 \\
                                 Slope & $-$0.70 $\pm$  0.13 &  $-$0.80 $\pm$  0.09 &   0.85 $\pm$  0.19 &  $-$2.19 $\pm$  0.08 &  $-$1.86 $\pm$  0.10 &  $-$1.02 $\pm$  0.03 \\
\enddata
\tablecomments{
Line fluxes and continuum flux densities at 5100 \AA\ ($F_{5100}$) are given in units of $10^{-18}$ erg s$^{-1}$ cm$^{-2}$ and $10^{-21}$ erg s$^{-1}$ cm$^{-2}$ \AA$^{-1}$, respectively. The flux upper bounds are defined as the 3$\sigma$ upper limits.
}
\end{deluxetable*}

\end{document}